# Implications of phonon anisotropy on thermal conductivity of fluorite oxides

Saqeeb Adnan[1], Miaomiao Jin[2], Matthew S Bryan[3], Michael E Manley[3] and David H. Hurley[4], Marat Khafizov[1]‡

[1] Department of Mechanical and Aerospace Engineering, The Ohio State University, Columbus, OH 43210, USA
[2] Department of Nuclear Engineering, The Pennsylvania State University, University Park, PA 16802, USA
[3] Oak Ridge National Laboratory, Oak Ridge, TN 37830, USA
[4] Idaho National Laboratory, Idaho Falls, ID 84315, USA

Fluorite oxides are attractive ionic compounds for a range of applications with critical thermal management requirements. In view of recent reports alluding to anisotropic thermal conductivity in this face-centered cubic crystalline systems, we perform a detailed analysis of the impact of direction-dependent phonon group velocities and lifetimes on the thermal transport of fluorite oxides. We demonstrate that the bulk thermal conductivity of this class of materials remains isotropic despite notable anisotropy in phonon lifetime and group velocity. However, breaking the symmetry of the phonon lifetime under external stimuli including boundary scattering present in nonequilibrium molecular dynamics simulations of finite size simulation cell gives rise to apparent thermal conductivity anisotropy. We observe that for accurate determination of thermal conductivity, it is important to consider phonon properties not only along high symmetry directions commonly measured in inelastic neutron or x-ray scattering experiments but also of those along lower symmetry. Our results suggests that certain low symmetry directions have a larger contribution to thermal conductivity compared to high symmetry ones.

## I. INTRODUCTION

First-principles understanding of phonon-mediated thermal transport in fluorite oxides with lanthanide and actinide elements as cations has been complicated by competing effects of lattice anharmonicity and electron correlation effects [1–3]. Analysis of harmonic and anharmonic properties depends on the precise computation of interatomic forces which strongly depends on the ability to accurately describe the electronic structure of the system [4,5]. While different methods to improve the electronic structure calculations in strongly electron correlated systems are available, the implications of these methods on the accuracy of thermal transport prediction are not known. Uranium dioxide ($UO_2$) is an example system where understanding thermal transport from first principles has been challenging [1,3,5]. In order to determine the adequacy of a simplified description of the correlation effects for a reliable prediction of phonon properties, it is important to understand various features of phonon dispersions and linewidths in these materials. These features include strong anharmonicity, high melting temperature, contribution of optical phonons to thermal transport, and emergence of resolvable Raman active vibration modes when defects are present [1,6–10]. Here, we focus on phonon wavevector direction dependent properties in thorium dioxide ($ThO_2$) as a model fluorite oxide system [6,11]. The acoustic anisotropy characterized by a directional dependence of the velocity of bulk plane acoustic waves in solids with cubic symmetries is well known [11,12]. Several recent reports have entertained a concept of thermal conductivity anisotropy in cubic fluorites such as $UO_2$, attributing their observations of phonon coupling to electronic degrees of freedom in strongly correlated systems [13–15]. The detailed characterization presented here of the phonon anisotropy in $ThO_2$ sheds light on this controversial topic, at the same time, provides a baseline for other fluorites, and assists in future studies focusing on the isolation of electron correlation effects on phonon structure and thermal transport.

$ThO_2$ is chosen as a model for fluorite oxides including $UO_2$, cerium dioxide ($CeO_2$), and yttria-stabilized zirconia (YSZ). $UO_2$ is the most extensively used commercial nuclear fuel [2], $CeO_2$ and YSZ have applications in the electrochemical energy conversion i.e., solid oxide fuel cell (SOFC) [16–18], while YSZ is also used as a thermal barrier coating. $ThO_2$ is also a surrogate for modeling $UO_2$ and $CeO_2$ properties using first-principles methods, as it offers an identical crystal structure of comparable phonon dispersions, but free from complications associated with the treatment of f-electrons correlation effects [19,20]. Its large optical bandgap makes it attractive as a laser host material. Most of these applications require a solid understanding of its thermal transport characteristics.

‡ khafizov.1@osu.edu

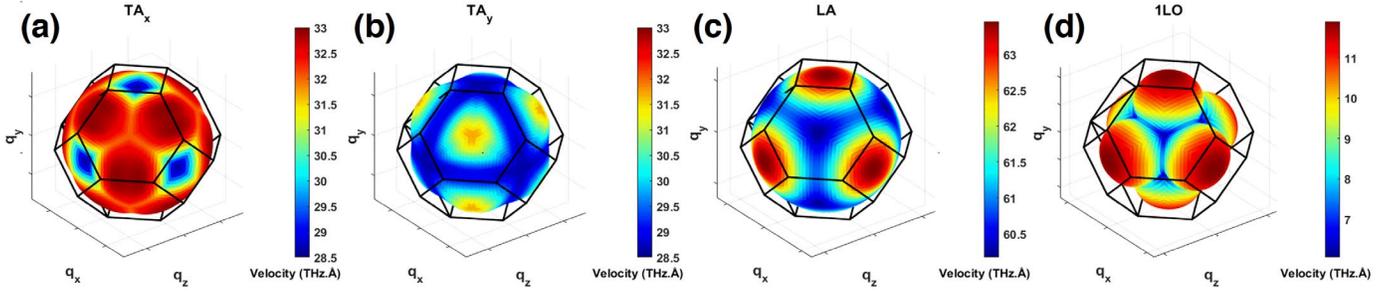

**FIG. 1.** 3D phonon velocity plots for $q = 0.1\frac{2\pi}{a}$ for acoustic (a) $TA_x$ (b) $TA_y$ (c) LA modes and optic (d) 1LO modes based on IFCs reported in [6].

Consequently, phonons and thermal transport in $ThO_2$ and related compounds have seen renewed interest. Phonon dispersion primarily determined by harmonic interactions has been investigated both experimentally [21] and theoretically [22,23]. At and above room temperature, lattice thermal conductivity is mainly limited by phonon-phonon scattering, whose strength is determined by lattice anharmonicity [24]. Measurement and modeling of anharmonic properties including phonon lifetimes have started to receive some attention [25,26]. Theoretical calculations of lattice thermal conductivity of $ThO_2$ using non-equilibrium molecular dynamics (NEMD) employing empirical interatomic potentials (EIP) [14,27,28], Slack method [29], and solution of Boltzmann transport equation (BTE) by using interatomic forces from ab-initio density functional theory (DFT) [6,22,30] have been reported. The impact of point defects, impurities, and grain boundaries resulting in additional phonon scattering in reducing thermal conductivity [31] have also been investigated [8,32,33]. A thorough discussion of these studies and several others can be found in a recent review article [2].

However, there still exists a debate over the anisotropic nature of thermal transport in fluorites. Despite having a cubic structure, Gofryk et al. [13] reported on experimental measurements indicating different thermal conductivity in $UO_2$ along its different crystallographic directions and attributed their observations to phonon scattering with electronic degrees of freedom whose strength depends on the direction of the applied temperature gradient. A computational study by Rahman et al. [14] using NEMD has also provided evidence of conductivity anisotropy in $ThO_2$ and $CeO_2$, acknowledging it is likely an artifact of finite simulation size. A recent study by Paolasini et al. [15] have reported on the direct measurement of anisotropy in phonon linewidth in $UO_2$ and suggests this could be the reason for observed anisotropic thermal conductivity. An accurate interpretation of the theoretical and experimental results requires a better understanding of its phonon properties. In our analysis of $ThO_2$, we found evidence of phonon anisotropy in the calculated phonon frequencies and phase velocities from the solution of the dynamical matrix using IFCs obtained from Jin et al [6,34]. Here, we define phonon anisotropy in the same spirit as acoustic anisotropy. Figure 1 depicts the surface plots of phonon phase velocity for selected phonon branches in $ThO_2$. The color bar represents the magnitude of the velocity of phonons propagating along different directions lying on the q-space iso-surface. The surface plots for acoustic phonons closely resemble the shape of the bulk acoustic velocities obtained from the solution of the Christoffel equation [12]. Acoustic modes exhibit a modest departure from the surface of the sphere, where LA (longitudinal acoustic) mode is close to directionally independent, while the TA (transverse acoustic) modes exhibit noticeable anisotropy. The velocity surface plot of the optical modes (transverse modes in supplementary document), on the other hand, exhibits complex structures having strong directional dependence and is thus classified as highly anisotropic.

Our classical understanding of lattice dynamics relies on measurements of phonons using inelastic neutron scattering experiments [1,21]. Early measurements employed nuclear reactor-derived neutrons utilizing triple-axis inelastic neutron scattering. This significantly limited the number of measurements and therefore has mostly been limited to measurement along high symmetry directions [21]. Novel instruments which utilize higher neutron fluxes derived from spallation sources such as wide-angle range time-of-flight neutron measurements allow measuring phonon dispersion over a broader range of phonon wavelength in a single acquisition [35]. While instruments utilizing wide angular range chopper spectrometer (ARCS) have been shown to be very effective in measuring phonon dispersion curves [36], isolating phonon lifetime from linewidths measurements is still complicated by an instrument resolution and requires further advancements before direct comparison to first-principles calculations is viable [25].

In this work, we investigate the impact of phonon anisotropy on lattice thermal conductivity within first-principles-based BTE formalism in $ThO_2$. It should be emphasized that we utilize well established methods for calculating thermal conductivity (Section II) and the unique aspect is in a careful analysis of the contribution of phonons propagating along different directions. The forthcoming discussion is organized into four sections. In section III, we demonstrate evidence of phonon anisotropy in $ThO_2$. At first, we compare two different approaches to compute lattice thermal conductivity. The first considers all the phonons, and the second represents the phonons using only high-symmetry directions. Such an approach helps us demonstrate the importance of considering phonons along low-symmetry directions. By analyzing the spectral conductivity profiles, we found a significant contribution from the off-axial phonons to thermal conductivity. To correlate these features with phonon properties, we carefully examine the phonon lifetimes and velocities in all the directions in q-space. Our results exhibit strong evidence of phonon anisotropy. However, we would like to emphasize that the bulk thermal conductivity in $ThO_2$ or other fluorites remains isotropic due to cubic symmetry as discussed in section IV [37]. Finally, in section V, we demonstrate how phonon anisotropy manifests itself in NEMD calculations, which is amplified as the size of simulation cells is reduced. These cell size-dependent results are found to be consistent with a simple BTE analysis including boundary scattering that captures and lifts the symmetry of the phonon lifetimes. Together these analyses provide a comprehensive insight into the implication of phonon anisotropy in $ThO_2$.

## II. METHODS

To calculate the thermal conductivity of $ThO_2$, we used Boltzmann Transport Equation (BTE) formalism under the relaxation time approximation as implemented within Phono3py code [38,39]. The conductivity tensor $k_{\alpha\beta}$ is computed using

$$k_{\alpha\beta} = \frac{1}{V_0} \sum_{\mathbf{q},m} C_{\mathbf{q},m} v_{\mathbf{q},m,\alpha} v_{\mathbf{q},m,\beta} \tau_{\mathbf{q},m} \qquad (1)$$

Here, $v_{\mathbf{q},m}$, $C_{\mathbf{q},m}$ and $\tau_{\mathbf{q},m}$ are the group velocity, mode heat capacity, and phonon relaxation time of the phonon mode, $m$, defined by the phonon wavenumber ($\mathbf{q}$) in reciprocal space. $V_0$ is the volume of a unit cell. Conductivity ($k_{\hat{n}}$) along a particular direction $\hat{n}$ can be calculated after replacing $v_\alpha v_\beta$ in Eq. 1 by $|v_{\mathbf{q},m} \cdot \hat{n}|^2$. In this formulation, $k$ includes the contribution from phonons propagating in all directions in the q-space. To emphasize the impact of phonon anisotropy, we introduce an alternative formulation that considers only phonons along high symmetry directions [00ξ], [ξ0ξ], [ξξξ]. This has been an approach taken when phonons could not be computed reliably, but experimentally available via traditional methods such as inelastic neutron scattering employing triple-axis spectrometer [21]. In this method, the whole Brillouin zone is approximated using the high-symmetry (axial) sampling points and the conductivity is given by [1]:

$$k_m^{(HS)} = \sum_{i=1}^{N_d} \Omega_i \int_0^{q_{max,i}} C_{q,m} \frac{v_{q,m}^2}{3} \tau_{q,m} q^2 dq \qquad (2)$$

Here, conductivity is calculated by the sum over selected directions in q-space namely the high symmetry phonons across three equivalent directions. Solid angle ($\Omega$) for each equivalent direction (6 along [00ξ], 12 – [ξ0ξ], and 8 – [ξξξ]) denoted by subscript $i$, are determined to be $\Omega_{[00\xi]} = 0.50$, $\Omega_{[\xi 0\xi]} = 0.58$, and $\Omega_{[\xi\xi\xi]} = 0.33$ and $N_d = 26$ [1]. The results using this formulation are termed 'high-symmetry' calculations from here on. Lastly, we use an expression applicable for isotropic phonons, where we consider only one direction in Eq. 2 with a solid angle of $\Omega = 4\pi$. The latter is reduced to a classical analytical expression [40].

To compute the parameters needed for thermal conductivity calculation including interatomic force constants (IFCs), we used phonopy [41] and phono3py [39] software packages. The second (2IFC) and third (3IFC) order IFCs were determined using DFT and CRG potential based on the finite displacement method with a displacement step of 0.03 Å following Jin *et. al* [6]. DFT calculations were implemented within VASP [42] where the electronic exchange correlation was described using local density approximation (LDA) [43]. The LAMMPS [44] package was used to calculate the forces of displaced supercell structure using an empirical interatomic force field based on CRG potential [45]. 2IFCs were used to construct a dynamical matrix for the calculation of phonon dispersions and 3IFCs were used to compute 3-phonon scattering rates. Results obtained from the DFT approach using the LDA will be denoted as DFT-LDA whereas results obtained using CRG potential will be denoted with EIP-CRG. The conventional unit cell consisting of 12 atoms is fully relaxed and their lattice constants are 5.529 A and 5.580 A for DFT-LDA and EIP-CRG, respectively. To calculate the 2IFCs, 3×3×3 supercells of conventional lattice consisting of 324 atoms are used, while 2×2×2 supercells of 96 atoms are used for 3IFCs. Further computational have been previously reported in [6]. Elastic constants calculated using these two approaches are $C_{11}$ = 352, $C_{12}$ = 113, $C_{44}$ = 72 GPa and $C_{11}$ = 380, $C_{12}$ = 130, $C_{44}$ = 96 GPa, for EIP-CRG and DFT-LDA, respectively. These values compare favorably with experimental data, reporting $C_{11}$ = 361, $C_{12}$ = 115, and $C_{44}$ = 78 GPa [26]. These values result in acoustic anisotropy represented by Zener ratios ($r$) of 0.602 and 0.768 for CRG and LDA, respectively. This indicates large anisotropy of phonon derived from EIP-CRG, which agrees better with the experimental value of 0.63. The EIP-CRG derived phonons are used as an illustrative example to amplify features derived from phonon anharmonicity.

The assignment of 9 phonon modes resulting from a 3-atom unit cell as $TA_x$, $TA_y$, LA, $1TO_x$, $1TO_y$, 1LO, $2TO_x$, $2TO_y$, and 2LO was performed based on the eigenvector analysis [46]. A and O refer to acoustic and optic modes, respectively. T and L correspond to modes having a primarily longitudinal and transverse displacement of atoms with respect to **q**. Subscripts x and y distinguish two transverse modes by their primary polarization along the x and y-axis when **q** is defined within irreducible Brillouin zone (BZ) confined by [001], [101], [111] directions. Eigenvectors of these modes contain information regarding the relative movements of the atoms i.e., the acoustic modes are primarily related to the vibration of heavier thorium ions whereas the optical modes are governed by the motion of oxygen ions. The mode assignment is done by first taking eigenvectors of the nine degenerate modes along [ξ0ξ] direction as a reference for each mode label. Then the projection of different phonons along those reference modes is calculated by using the dot product of the phonon's mode eigenvector and each reference. The mode with the largest projection is assigned to that reference mode.

Inelastic neutron scattering (INS) measurements were performed to experimentally measure the phonon dispersion curves of $ThO_2$. Time-of-flight inelastic neutron scattering measurements were performed on a single crystal of $ThO_2$ at 300 K, as described elsewhere [35] using the wide Angular Range Chopper Spectrometer (ARCS) [36]. A pulse of monochromatic incoming neutrons exchanges energy and momentum with the sample. Position sensitive detectors measure the scattered neutron intensity as a function of angle and time, which is then converted into scattered intensity as a function of momentum and energy transfer. Multiple Brillouin zones are contained in the full data set and scattering as a function of energy is analyzed along the high symmetry directions.

The phonon dispersion and linewidths measured with INS can in principle be used to determine $v_{\mathbf{q},m}$ and $\tau_{\mathbf{q},m}$, respectively. The results obtained from ARCS can be readily processed to construct phonon dispersions and direct comparisons to model results. However, for phonon linewidths, which are inversely proportional to phonon lifetime, the comparison is not straightforward, and one needs to account for instrumental broadening. The INS measurements of $ThO_2$ contain the relaxation time of a small volume in reciprocal space, in contrast to the points in reciprocal space that are produced by simulation. The volume size is adjustable during data analysis but limited by instrumentation, and sample size among other factors. There is an appreciable change in relaxation time across the volume measured by INS, and as a result, the INS relaxation times are an average over the measured volume. This is in contrast to simulation results which occur at a single point in reciprocal space, and a direct comparison does not yield an agreement. Simulating a volume identical to the measured INS volume is required. The phonon dispersion does not appear to be sensitive to this effect. A detailed discussion of this subject is beyond the scope of this manuscript [25].

Non-equilibrium molecular dynamics (NEMD) as implemented in the LAMMPS [44] package is also used to calculate the thermal conductivity along different crystallographic directions in $ThO_2$ [32]. The interatomic interactions are described by the CRG potential [45]. An elongated simulation cell having periodic boundary conditions along all directions is constructed as two regions are set to be hot and cold zones, respectively. The temperature of the two regions is controlled via the Langevin thermostat. The thermal conductivity is calculated using Fourier's law after a stable linear temperature profile and constant heat flux are established. For this study, data are collected over 1 ns long simulation employing the microcanonical ensemble with a time step of 0.001 ps. The length of the simulation cell is varied to capture the size effect on thermal conductivity. For each simulation cell length, three sets of calculations are performed with different initializing random seeds in order to improve statistical accuracy.

## III. PHONON ANISOTROPY

### A. Phonon dispersion

We start by analyzing the phonon dispersion curves for $ThO_2$. In Fig. 2, we present the experimental results obtained from inelastic neutron scattering compared to the DFT-LDA results. Experimental dispersion for high symmetry directions along Γ–X [ξ00], Γ–K [ξξ0], and Γ–L [ξξξ] have been previously reported [35], whereas the X-W direction is reported for the first time. Model dispersion curves include additional low symmetry directions such as Γ–W [2ξ,ξ,0] and X-L. It should be noted here that in fact Γ–W direction has higher symmetry than Γ–K, but in this manuscript, we still refer to Γ–W as low symmetry, to keep it consistent with the historic trend where Γ–K has been more frequently reported.

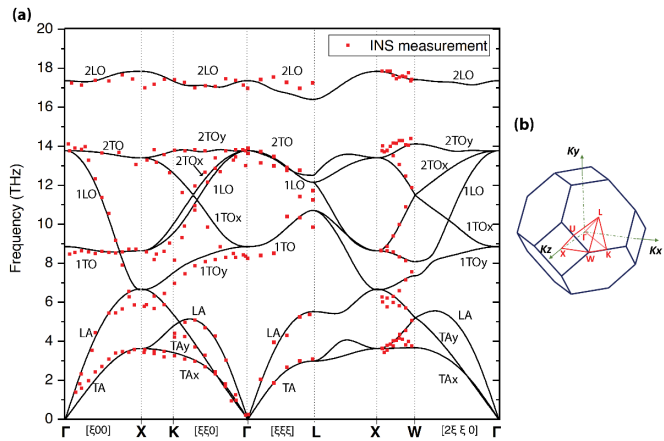

**FIG. 2.** *(a) Phonon dispersion curves for $ThO_2$ using DFT-LDA and experimental results from INS. (b) First Brillouin zone and the irreducible Brillouin zone wedge (red).*

The slopes of Γ–X, Γ–K, Γ–L, and Γ–W define the group velocity, from which an initial assessment of the relative contribution of each phonon mode to thermal conductivity can be performed. $ThO_2$ dispersion is represented by nine modes (three acoustic and six optical modes) corresponding to 9 degrees of freedom of the three atoms in the primitive cell [47]. The acoustic modes are classified into two transverse acoustic (TA) and longitudinal acoustic (LA) modes. Along Γ–X and Γ–L directions, the two transverse modes are degenerate, but this degeneracy is lifted along other directions. We distinguish TA modes based on the direction of atomic vibration and term them as $TA_x$ and $TA_y$ (polarized along the x and y-axis respectively for modes inside the irreducible BZ shown in Fig. 2b). Careful examination of dispersion curves reveals a strong anisotropy in $TA_y$ mode. While LA mode branches along high symmetry directions overlap, i.e., they all have maxima at about 5 THz, the $TA_y$ mode's branches are spread out where maximum frequency varies over 3 - 6 THz range along different directions. Acoustic modes have large slopes as generally expected and should contribute significantly to the conductivity of $ThO_2$ [1,48–50]. Fluorite structures possess highly dispersive optical modes. It can be seen that the LO mode has a high group velocity along the Γ–X and Γ–K directions and its contribution to thermal transport has been discussed in several recent reports [1,3,7]. However, velocity along Γ–L direction is smaller. This highly anisotropic nature of optical modes is best captured in Fig. 1(d). It should also be noted that $1TO_y$ while relatively flat along high symmetry directions, exhibits a large slope along the low-symmetry directions (supplementary document).

A closer examination of dispersion along the X-W segment serves as illustrative experimental evidence of strong anisotropy in phonon structure. It approximates an arc on the sphere defined by constant q-magnitude. For isotropic material, the frequency of individual modes should remain constant. A large variation of phonon frequencies along this path validated by INS measurements is a strong indication of phonon anisotropy in this fluorite system. We point out that this anisotropy in not unique to fluorites, as other cubic materials like diamond also have dispersion profiles with similar anisotropy along high symmetry directions whereas in silicon carbide the anisotropy is less obvious [51,52]. To perform a more comparative analysis, instead of using a different material, we consider $ThO_2$ phonons as predicted by an alternative force field based on EIP-CRG. Dispersion curves based on IFC derived from EIP-CRG have comparable acoustic mode features, but optical modes are notably different [6]. This serves as a primary reason to include EIP-CRG as a model that exhibits a different degree of anisotropy.

### B. Branch resolved high symmetry conductivity

The mode-specific thermal conductivity contributions calculated using DFT-LDA and EIP-CRG derived IFCs are depicted in a bar plot in Fig. 3. To demonstrate the impact of phonon anisotropy we provide conductivities calculated using two methods. The results presented in red consider phonons along all directions, while results in blue consider only high symmetry modes in keeping with historic measurements [1,21,35]. The total thermal conductivity found using the high symmetry method is 14.20 W/m·K, compared to 17.96 W/m·K using the all-phonon method in the case of LDA calculation. A similar calculation using EIP-CRG results in 16.64 vs 23.61 W/m·K – a much larger difference. The high symmetry method notably underestimates thermal conductivity. When one considers

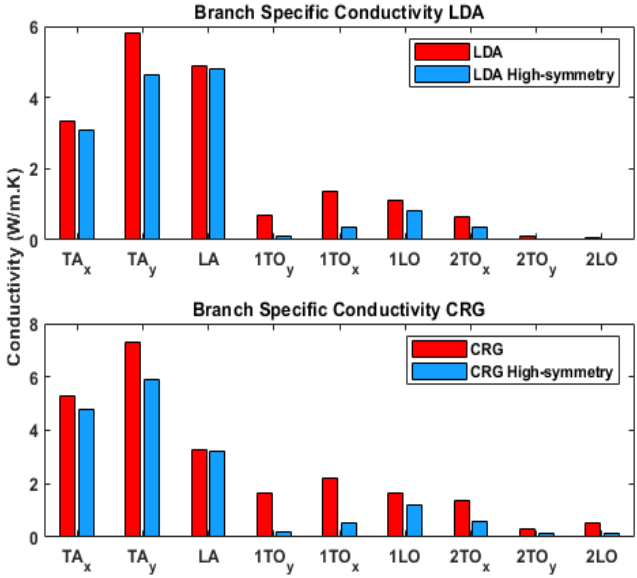

*FIG. 3. Mode-specific thermal conductivity contribution for DFT-LDA and EIP-CRG using all-phonon (red bars and Eq.1) and high-symmetry phonon (blue bars and Eq.2) methods*

mode-specific conductivity, one observes higher values from the first method compared to the high-symmetry approach for most of the modes. While both approaches provide a comparable contribution from LA modes, the discrepancy is the largest for $TA_y$ and all TO modes, which suggests that the low-symmetry direction phonons might provide a greater contribution than the high symmetry ones. Relative contribution trends for different modes are in agreement with the trends in phonon's group velocity revealed from an analysis of dispersion curves discussed in Sec. III. A and in Fig. 2, and have been addressed in previous reports [1,6].

## C. Spectral conductivity

The origin of the above-mentioned difference in thermal conductivity calculation can be inferred from careful analysis of a direction-dependent spectral conductivity for the optical and acoustic modes shown in Figs. 4-5. The spectral conductivity calculation for a particular direction follows the procedure of high-symmetry calculations, but in this case, only the phonons along a single direction are used to calculate the total conductivity, essentially limiting the summation in Eq. 2 to one direction with a $4\pi$ solid angle, resulting in:

$$\frac{dk}{dq} = 4\pi C_{q,m} \frac{v_{q,m}^2}{3} \tau_{q,m} q^2 \qquad (3)$$

Figure 4 shows the spectral conductivity profiles for three of the optical modes, namely 1LO and two 1TO modes, exhibiting the largest degree of anisotropy. There is a noticeable directional dependence for 1LO mode, where [ξξξ] direction has a very little contribution, whereas [00ξ] direction has a very large contribution (Fig.4c). Similar anisotropic features are found in the $1TO_x$ mode (Fig.4a); however, one feature stands out for $1TO_x$ mode. While low symmetry directions in 1LO are mostly bounded by the high symmetry ones, the [2ξ,ξ,0] direction of $1TO_x$ mode is outside of the bounds defined by the high symmetry axis. Note that the other low symmetry direction phonons are represented by the scatter points. Considering this together with results in Fig. 3, strongly suggests that low symmetry directions contribute significantly more to conductivity than high symmetry phonons for $ThO_2$. The spread of conductivity values for a fixed q-value can be used as a qualitative indication of the degree of anisotropy. While anisotropic features are most notable in the optical modes, their low contribution to thermal conductivity alone makes a weak impact on the total conductivity.

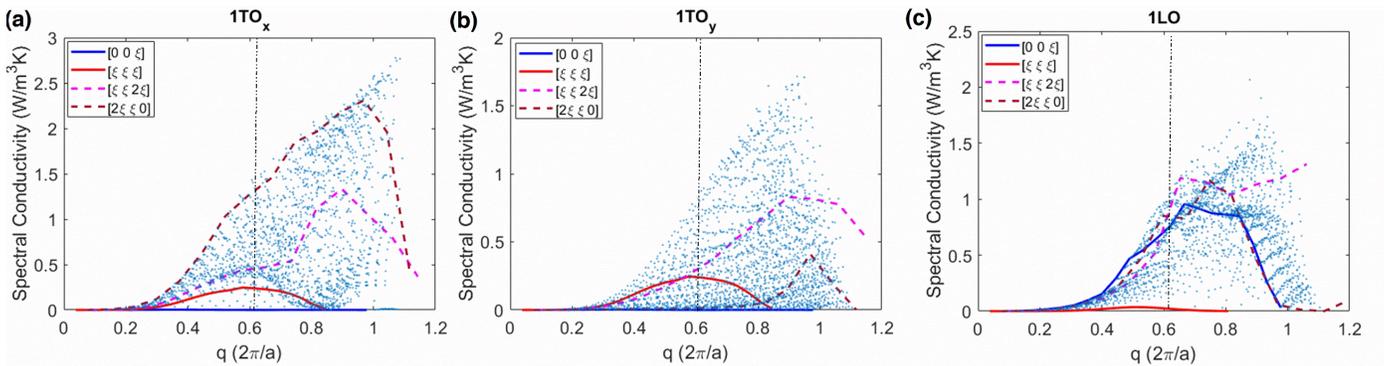

*FIG. 4. Spectral conductivity profiles for optical modes (a) $1TO_x$ (b) $1TO_y$ and (c) 1LO using DFT-LDA. Spectral conductivity values are calculated by multiplying the thermal conductivity values with a weighting factor based on the q-point and phonon orientation.*

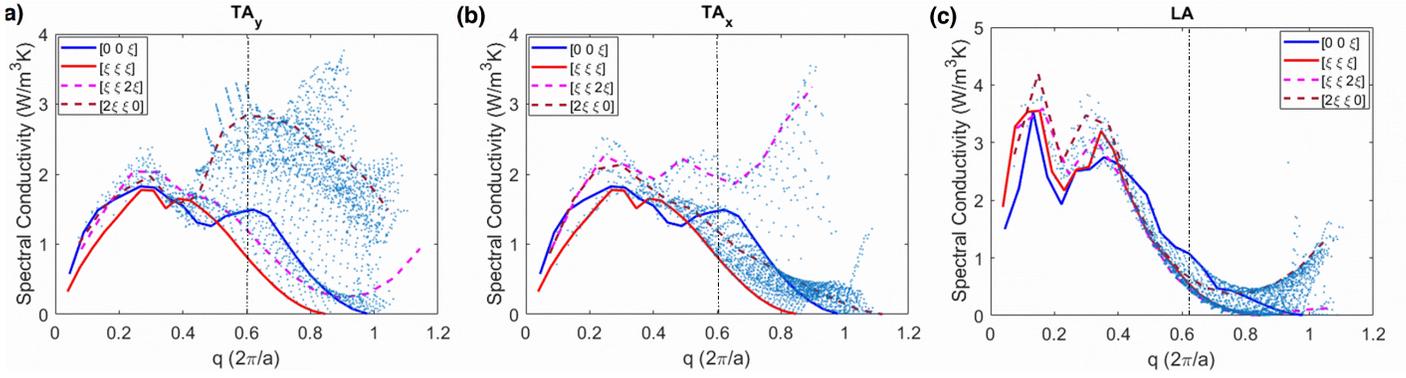

**FIG. 5.** Spectral conductivity profiles for acoustic modes (a) $TA_x$ (b) $TA_y$ and (c) LA using DFT-LDA.

Acoustic modes have similar anisotropic features as shown in Fig. 5. Anisotropic features in the spectral conductivity profiles are more evident for $\frac{qa}{2\pi} > \frac{1}{2}$. The spectral conductivity profiles for $TA_y$ mode (Fig. 5a) are highly anisotropic and suggest significant contributions from the phonons along low symmetry directions. The spectral conductivity profile for $TA_x$ mode (Fig. 5b) also indicates a higher contribution from a few of the phonons along a low symmetry direction. Similar to the case of $TO_x$ mode, these phonons i.e., one along [ξ,ξ,2ξ] direction, are not bounded by the high symmetry axes. In contrast, these anisotropic features are less obvious in the LA mode (Fig. 5c). In fact, one can classify the LA mode as isotropic. There is also a dip present in the spectral conductivity profile along the [00ξ] direction for all acoustic modes, located at 0.2 for LA and 0.45 for TA modes.

The spectral conductivity anisotropy is stronger in the model system represented by EIP-CRG, an empirical interatomic potential, exhibiting a stronger anisotropy.

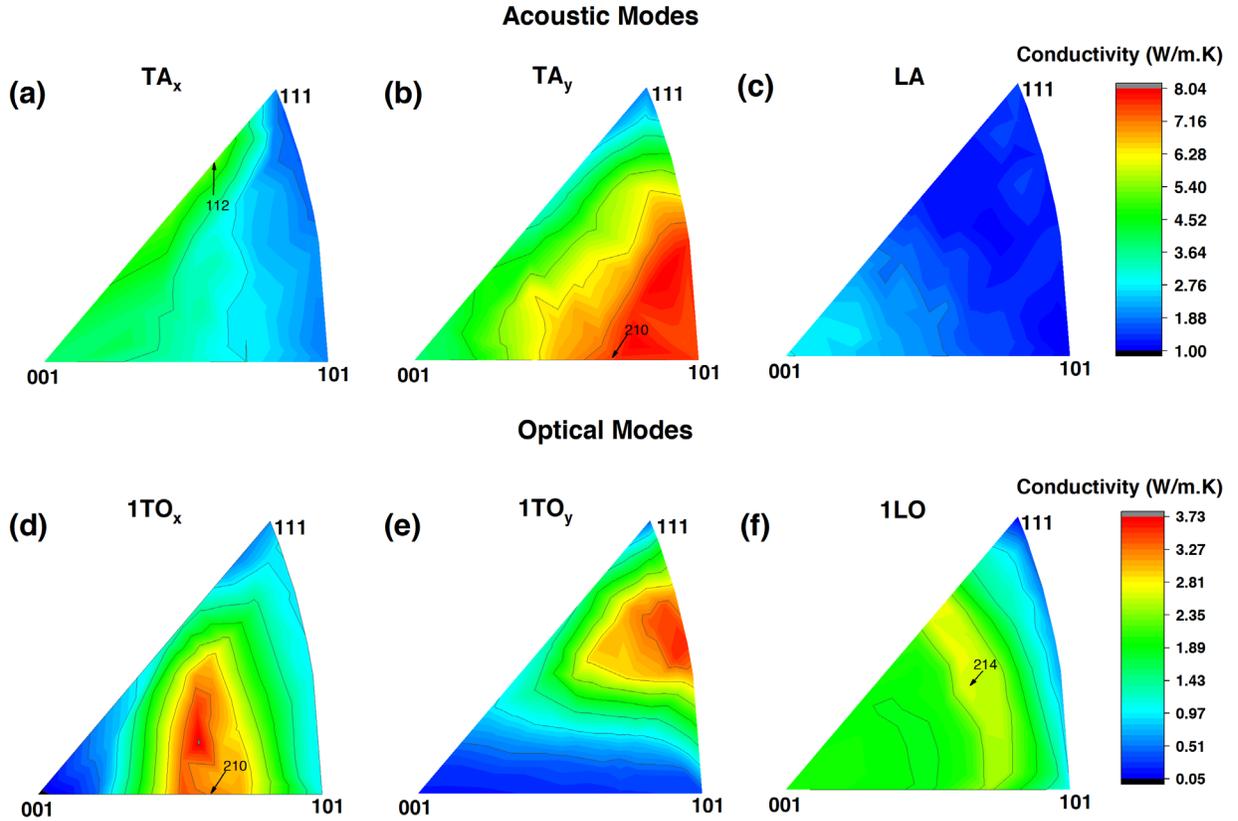

**FIG. 6.** Inverse polar plots of the acoustic modes (a) $TA_x$, (b) $TA_y$, (c) LA, and optical modes (d) $1TO_x$, (e) $1TO_y$, (f) $1LO$ using DFT-LDA at q=0.620. Arrows indicate particular modes discussed in the text.

Figure S7 (in supplementary document) depicts the spectral thermal conductivity profile for the acoustic modes calculated using the CRG potential. Similar to the case of DFT-LDA calculation, the TA modes have large thermal conductivity contributions from phonons along low symmetry directions (Fig. S6 a and b). However, with EIP-CRG, the anisotropic features in the LA mode are more pronounced over q-values ranging from $0.2 < \frac{qa}{2\pi} < 0.4$. Therefore EIP-CRG derived phonon properties can serve as a model for higher anisotropy system. It is to be noted that the contribution of optical branches to thermal conductivity is much more significant with EIP-CRG compared to DFT-LDA as can be seen in Fig. 3. The spectral conductivity profiles for acoustic modes along [00ξ] direction have a dip similar to the DFT-LDA case.

An alternative method to represent the anisotropic feature is to consider the contribution of all modes propagating in all-direction using a contour plot similar to the q-iso-surface plots shown in Fig. 1. For this, we chose irreducible representation using a 2D surface plot, where we only plot the wavevectors contained within the irreducible BZ shown in Fig. 2(b) corresponding to a spherical shell with a radius $\frac{qa}{2\pi} = 0.62$. From Fig. 6, it can be observed that low symmetry directions optical modes represented by the central portions of the polar plots have significantly greater contributions compared to the high symmetry phonons located at the corners. This is especially notable in the case of 1TO$_x$ mode (Fig. 6d), which has maxima along [hkl] = [210]. On the other hand, for the 1LO (Fig. 6f), the anisotropy effect is weak. As for the acoustic branches, LA mode appears isotropic consistent with Fig. 5(c). In TA modes, for the most part, the contribution of low-symmetry acoustic modes is bounded by the high-symmetry phonons. Similar polar plots for EIP-CRG calculation are available in supplementary documents and unlike the DFT-LDA case, it reveals strong anisotropy within LA mode.

### D. Analysis by phonon group velocity and lifetime

Here, we discuss the source of the anisotropic thermal conductivity contributions of different modes in ThO$_2$. The spectral conductivity profiles and polar plots are presented in Figs. 4-6 illustrate the significant contribution of low-symmetry phonons to thermal conductivity. Here, we take a closer look at the phonon velocity and lifetimes as both contribute to the calculation of thermal conductivity (Eq. 1). Figure 7 shows the phonon velocities and lifetimes for selected acoustic and optical modes obtained from DFT-LDA IFC (plots for all 9 modes are available in the supplementary document). It should be emphasized that comparison should be done on modes that have the same q-vector magnitude. In the case of the TA$_x$, some modes are seen to have higher velocity but shorter lifetimes than the high-symmetry modes. In particular, [ξ,ξ,2ξ] mode appears to have the largest velocity, which also has the largest contribution to the spectral conductivity profiles (Fig. 5b). In the case of TA$_y$ and LA modes, we observed that the phonon velocities and lifetime profiles for off-axial phonons are bounded by the high-symmetry ones for the most of directions. One interesting observation is that the conductivity contribution of [00ξ] and [ξξξ] phonons vanish at the zone boundary. However, the contribution of lower symmetry phonons such as [ξ0ξ] and [2ξ, ξ,0] are large at zone boundaries. All these zone boundary effects correlate with the magnitude of the velocity (Fig. 7). This results in high spectral conductivity for some of the low-symmetry phonons at high q-magnitude (Fig. 5).

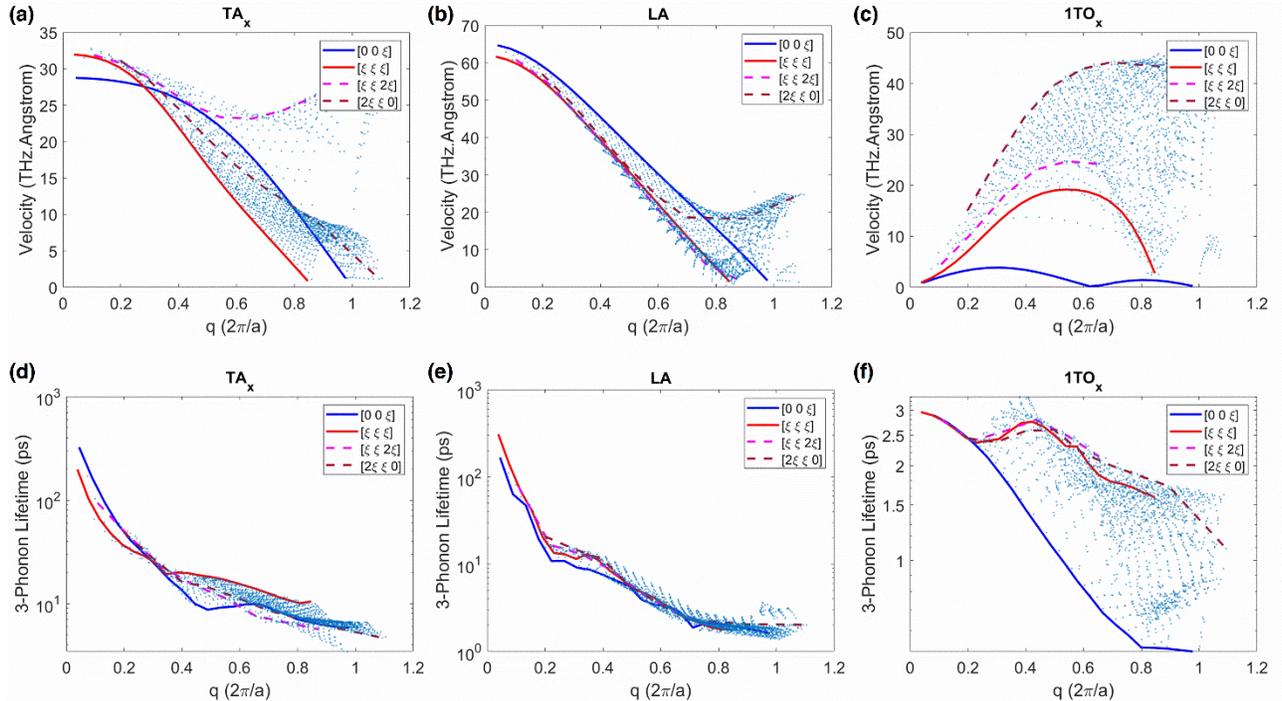

***FIG. 7.*** *Phonon group velocities (top) and lifetimes (bottom) for TA$_x$, LA, and 1TO$_x$ mode from DFT-LDA.*

In the case of the optical modes, we have seen a high level of anisotropy in the $1TO_x$ mode (Fig.4a). Examination of plots in Fig. 7(c) reveals that the velocity for low symmetry phonon is significantly larger than for high-symmetry ones. Another notable observation, unlike acoustic modes, the distribution of the $TO_x$ mode's lifetime is not bounded by high-symmetry phonons. The combined effect of velocity and lifetime results in large anisotropy in spectral conductivity profiles. Comparing phonons along [ξξξ] and [2ξ,ξ,0] directions, one can observe that they have comparable lifetimes. However, the ones along [2ξ,ξ,0] are significantly faster, resulting in a very high spectral conductivity. Thus, combining the results for both acoustic and optical modes (Fig. 7), a larger phonon velocity along low-symmetry directions can be attributed as the primary factor responsible for the observed anisotropic behavior in spectral conductivity.

As already pointed out, the spectral conductivity plots show a consistent dip in the spectral profiles of the acoustic modes, especially along [00ξ] direction appearing in $\frac{qa}{2\pi} = 0.38 - 0.52$ range for both TA modes, which corresponds to a frequency range of 2~2.5 THz. We speculate that this dip is related to both the magnitude of the three-phonon scattering phase space and scattering strength. The former is quantified by the joint density of states (JDOS) $D_2$ [53]. We noticed a sharp rise in the scattering strength of the phonons for the TA mode located in the vicinity of the dip (see Fig. S5). Both the scattering strength and phase space lower the thermal conductivity by enhancing the 3-phonon scattering rate [31,54].

Anisotropic features observed in the acoustic branches have practical implications. Elastic anisotropy has been previously utilized to measure elastic constants in irregularly shaped single crystals, image grain boundaries and measure the orientation of the crystal surface in polycrystalline materials [12,26,55,56]. The orientation-dependent velocity profile of the crystals has been used to gain information on the grain orientation. A similar method can be implemented in the case of thermal anisotropy to capture the microstructural information of the material. However, unlike elastic anisotropy which can be represented by a single scalar term Zener ratio [57], thermal anisotropy has much more complexity as it is sensitive to several other parameters like phonon frequency and q-point vector which leaves scope for future investigation.

## IV. ISOTROPIC BULK CONDUCTIVITY

Recent experimental study using PPMS (Physical Property Measurement System) to measure thermal transport in $UO_2$ has reported a controversial observation suggesting their results provide evidence of anisotropic thermal conductivity in $UO_2$ [13]. However, it should be emphasized that despite the anisotropic contribution of different modes, the total bulk conductivity is isotropic as expected for materials with cubic symmetry. This is because thermal conductivity is a second-rank tensor and any physical property having a second-rank tensor will always be isotropic for a cubic material [58]. In the appendix, we provide analytical proof that the conductivity calculation using Eq. 1 indeed results in isotropic conductivity and demonstrate that this isotropy is dictated by the symmetry of the $\mathbf{v} \otimes \mathbf{v}$ product. Mathematically, this leads to a diagonal conductivity tensor having a single independent component which indicates isotropic thermal conductivity. It should also be noted that the aforementioned report attributed larger conductivity along [110] and [111] than [100] directions to an emergence of additional phonon scattering that depends on the heat flux direction. This implies that some nonlinear behavior and a reduction in crystal symmetry under applied heat flux may serve as a plausible explanation for the apparent thermal conductivity anisotropy. This nonlinear behavior had not been fully explored, and the inherent limitation of the PPMS measurements leaves this apparent anisotropy an open question.

An experimental evidence for phonon linewidth anisotropy in $UO_2$ has been recently reported by Paolasini et. al [15] and had been used as potential explanation for the apparent thermal conductivity anisotropy in $UO_2$. However, it is important to point out that our results demonstrate that the phonon linewidth anisotropy is not an indication of conductivity anisotropy. Despite phonon linewidth being directly related to phonon relaxation time which contributes to thermal conductivity, anisotropic broadening of phonon linewidth does not result in anisotropic conductivity (Appendix A). Therefore, we rule out Paolasini's observations as a possible explanation of apparent thermal conductivity anisotropy reported in [13]. The following section aims to explain the recently reported apparent anisotropic thermal conductivity in fluorite oxides under the conditions when the cubic symmetry is broken.

## V. BOUNDARY SCATTERING INDUCED ANISOTROPY

So far, our discussion has focused on the anisotropic nature of the phonons, and based on the discussion in the previous section, we concluded that the bulk conductivity in ThO$_2$ and other fluorites is isotropic. However, there are special cases where the strong phonon anisotropy can lead to an induced conductivity anisotropy. It is expected to occur when some external perturbation leads to a reduction in the symmetry of the phonon lifetime. One example involves the thermal conductivity calculation using NEMD. Figure 8 shows the computationally measured thermal conductivity of ThO$_2$ as a function of simulation cell length and crystallographic orientation of the cells with respect to heat flux. We observe, not only that the thermal conductivity depends on the cell size but also on crystal orientation. Notably, anisotropy is negligible for large cells where all size effects vanish as expected. This size dependence has been attributed to ballistic effects, where the phonons with mean free path (MFP) larger than the cell size do not contribute to thermal conductivity [59]. We observe that for [110] and [111] orientations, the conductivity reduction is smaller compared to [100]. Notably this trend agrees with experimental observations accompanied by NEMD calculations reported in [13].

We analyze this behavior using BTE with the inclusion of a boundary scattering term. For comparative analysis with the NEMD results, we included a boundary scattering term in our BTE calculation. The total scattering time, $\tau$ was expressed using Matthiessen's rule as [31,60,61]:

$$\frac{1}{\tau} = \frac{1}{\tau_{3ph}} + \frac{2\vec{v} \cdot \hat{h}}{L} \quad (4)$$

Here, $\tau_{3ph}$ is the three-photon scattering time obtained from Phono3py and 2$^{nd}$ term is the boundary scattering time. The boundary scattering time is expressed in terms of the simulation cell length ($L$) and the inner product of the phonon velocity ($\vec{v}$) and the directional vector ($\hat{h}$). This direction specifies crystalline orientation with respect to the boundaries imposed by the finite length of the NEMD cell along the heat flux direction.

We noticed that BTE conductivity values in the limit of large cells are smaller. Several factors can contribute to this. First, BTE is not capturing phonon interactions resulting from beyond 3$^{rd}$-order anharmonicity, while NEMD does [62]. Additionally, phonon distribution in NEMD lacks quantum effects present in BTE. We also noticed that the NEMD results showed a significant level of anisotropy compared to the BTE, this could also be related to the same factors that lead to a discrepancy in the bulk values. Next, regarding the actual origin of the anisotropic conductivity, we attribute this to the length of the simulation cell being comparable to the mean free path (MFP) of the phonons. We see that the simple boundary scattering expression used in BTE is unable to capture the apparent anisotropy to the same extent as NEMD does. Nevertheless, the trends are similar which suggests that size effects impact [001]-oriented cells more than [101] and [111]-oriented ones. A simple explanation for this can be obtained by examining the spectral conductivity plots (Fig. 5), particularly focusing on low q-value phonons that have the largest MFP and are most impacted by size effects. For the acoustic modes at low q-values, the contribution of the [001]-phonons are larger than [101] or [111] ones (Fig.5). In small cell sizes, this impacts [001]-phonons the most, as a result leading to the larger reduction in conductivity for cells oriented along this direction. Similar observations can be made from the 3D velocity plots in Fig. 1 for the low value of $\frac{qa}{2\pi} = 0.1$. For both the transverse acoustic modes, phonons have low velocity in the [001] direction compared to the [101] and [111] directions. That means the phonons will scatter heavily in [101] and [111] directions, whereas a weak scattering is observed in the [001], resulting in a relatively high contribution for [001]-phonons [63]. The scope of this anisotropic behavior is limited to the contribution of off-axial phonons towards conductivity calculated inside a finite simulation cell. If the size effect of the simulation cell can be eliminated, in other words, if infinite cell length is assumed as for bulk material, the thermal conductivity remains the same along all directions.

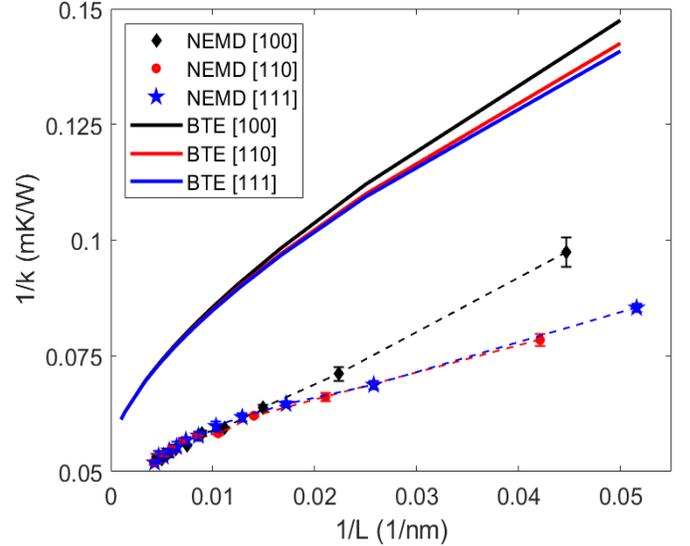

**FIG. 8.** *Thermal conductivity as a function of crystal orientation and simulation cell length using NEMD and BTE.*

These observations emphasize the importance of size-dependent analysis when using NEMD. These same arguments provide an explanation for the recent reports of anisotropic conductivity in fluorite oxides based on the result of NEMD calculations [7,13,14]. The above demonstration can be expanded to include other external factors and where inclusion of other mechanisms might provide a viable explanation for experimentally observed apparent thermal conductivity reported in [13].

## VI. CONCLUSIONS

In this work, we have calculated the lattice thermal conductivity of ThO$_2$ based on first-principles calculation (DFT) using two contrasting approaches. One of them considers only the phonons along high-symmetry directions while the other considers all the phonons to calculate conductivity. The spectral conductivity profiles and iso-surface plots showed strong signs of phonon anisotropy, particularly in the TA$_y$ and the optical modes. Although the bulk conductivity in ThO$_2$ remains isotropic due to O$_h$ symmetry, the size-dependent NEMD results presented a special case where the thermal conductivity was anisotropic. Similar trends were observed with the inclusion of boundary scattering terms in BTE calculation which lifted the cubic symmetry of phonon lifetimes. These cases demonstrated how phonon anisotropy may lead to apparent anisotropic thermal conductivity which has been reported recently for UO$_2$. The comprehensive analysis of the phonon anisotropy in ThO$_2$ presented here, can provide better insight for accurately treating the intricacy involved in characterizing thermal transport in fluorites with strongly correlated electrons. Further exploration of phonon anisotropy in fluorites can aim at developing non-destructive surface imaging of crystals by utilizing orientation-dependent phonon properties. Focusing on the high energy carrying phonons one can also develop materials with a tailored thermal conductivity that can be utilized in nano and micro-scale energy applications.

## ACKNOWLEDGEMENTS

This work was supported by Center for Thermal Energy Transport under Irradiation (TETI), an Energy Frontier Research Center funded by the U.S. Department of Energy, Office of Science, and Office of Basic Energy Sciences.

## APPENDIX: PROOF OF ANISOTROPIC CONDUCTIVITY IN FLUORITE OXIDES

Fluorite has an *Fm3m* structure with O$_h$ symmetry of a face-centered cubic crystal whose Brillouin zone is represented by a truncated octahedron. Phonon properties can be evaluated within the irreducible Brillouin zone (IBZ) defined by the shaded wedge in Fig. 2. To evaluate the full BZ one applies 48 transformations $\boldsymbol{R}_i$ needed to populate the full BZ. When only intrinsic 3-phonon scattering exists, the system retains the O$_h$ symmetry.

If one is interested in conductivity along a particular direction $\hat{n}$, it is evaluated as a summation over each point in the BZ:

$$\kappa_{\hat{n}} = \frac{1}{V_0} \sum_{q \in BZ} C_q \tau_q |\hat{n} \cdot \vec{v}_q|^2, \quad (A1)$$

Where $\vec{v}_q$, $\tau_q$, and $C_q$ are the phonon group velocity, lifetime, and heat capacity for a point in the q-space inside the BZ. This can be rewritten in a form efficient for computation of bulk properties by considering only phonons within IBZ subjected to a condition that $\tau_q$ retains the O$_h$ symmetry:

$$\kappa_{\hat{n}} = \frac{1}{V_0} \sum_{q \in IBZ} C_q \tau_q \sum_{r=\{1:48\}} |n_\alpha R_{r,\alpha\beta} v_\beta|^2, \quad (A2)$$

where, $R_{r,\alpha\beta}$ are the 48 rotation matrices used to reproduce full truncated octahedron BZ from the irreducible wedge [64]. The inner sum in Eq. A2 can be evaluated analytically to show that:

$$\sum_{r=\{1:48\}} |n_\alpha R_{r,\alpha\beta} v_\beta|^2 = 16(v_x^2 + v_y^2 + v_z^2) \quad (A3)$$

This depends only on the magnitude of the group velocity and is independent of the direction along which conductivity is calculated. This indicates that even though phonons have strong anisotropy their impact on thermal conductivity is isotropic and is attributed to the symmetry of the FCC structure. Alternatively, one can calculate the conductivity tensor and show that a diagonal matrix with only a single independent component remains, consistent with Neumann's principle.

$$\sum_{r=\{1:48\}} R_{\alpha\gamma} v_\gamma R_{\beta\delta} v_\delta = 16(v_x^2 + v_y^2 + v_z^2) \begin{bmatrix} 1 & 0 & 0 \\ 0 & 1 & 0 \\ 0 & 0 & 1 \end{bmatrix} \quad (A4)$$

When symmetry-breaking phonon scattering processes are present, such as in the case of scattering from the boundaries, the simplification present in Eq. A2 is no longer applicable. If one accounts for the boundary scattering assuming that all of the boundaries are oriented with a surface normal $\hat{b}$, the scattering rate can be expressed as,

$$\frac{1}{\tau_B} = \frac{2|\vec{v}\cdot\hat{b}|}{L} \quad (A5)$$

This lift the symmetry for the scattering rate and the inner summation in the previous expression (Eq. A2) can no longer be performed with $\tau_q$ as a common factor. However, $\tau_q$ is expected to have the symmetry of $\hat{b}$ directions, such as (100), (110), and (111) will have $C_{4v}$, $C_{2v}$, and $C_{3v}$ symmetries respectively.

# Supplementary Material

This is a supplementary document with the additional plots for spectral conductivity, phonon lifetime, velocity, and frequency for all 9 phonon branches. Effect of isotopic scattering on the acoustic and optical modes are also present.

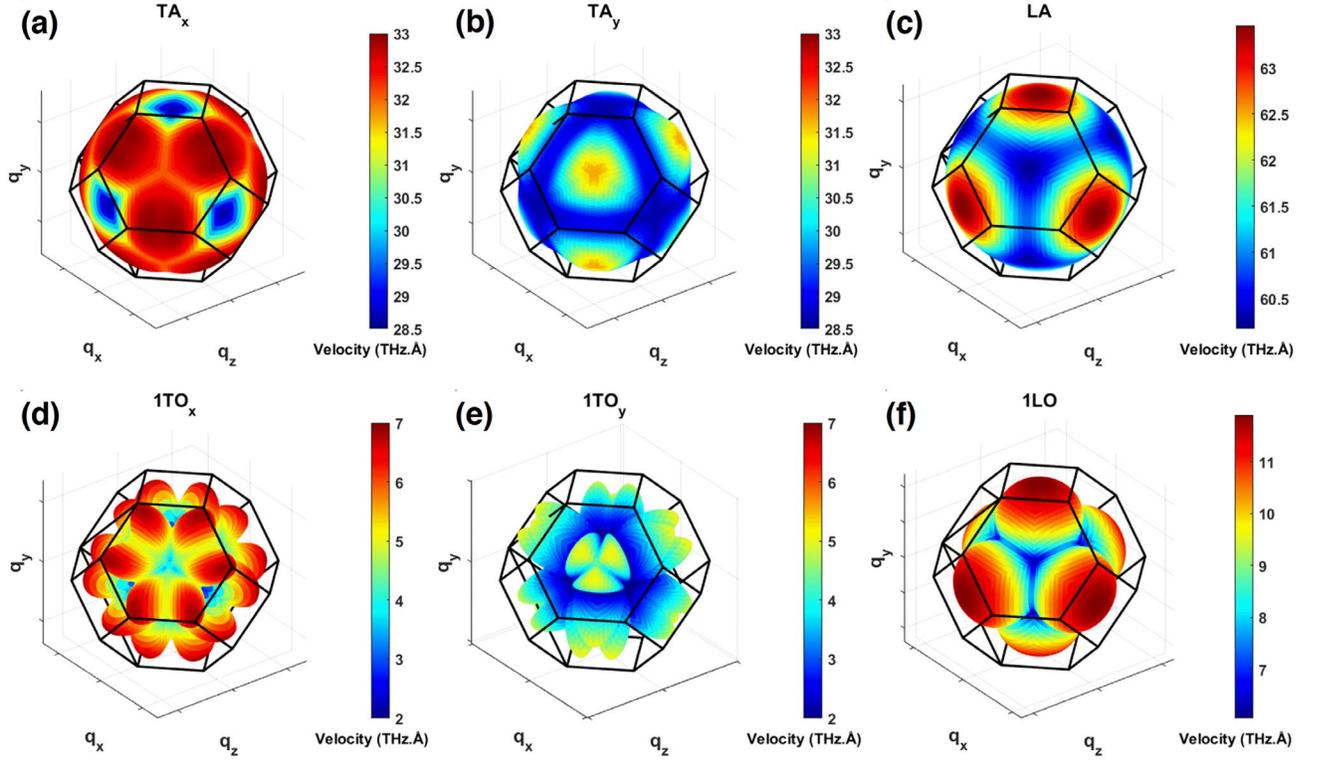

**FIG. S1.** *3D phonon velocity plots for q=0.1 2π/a for acoustic (a) TAx (b) TAy (c) LA modes and optic (d) 1LO modes based on IFCs reported in [6] in the main text.*

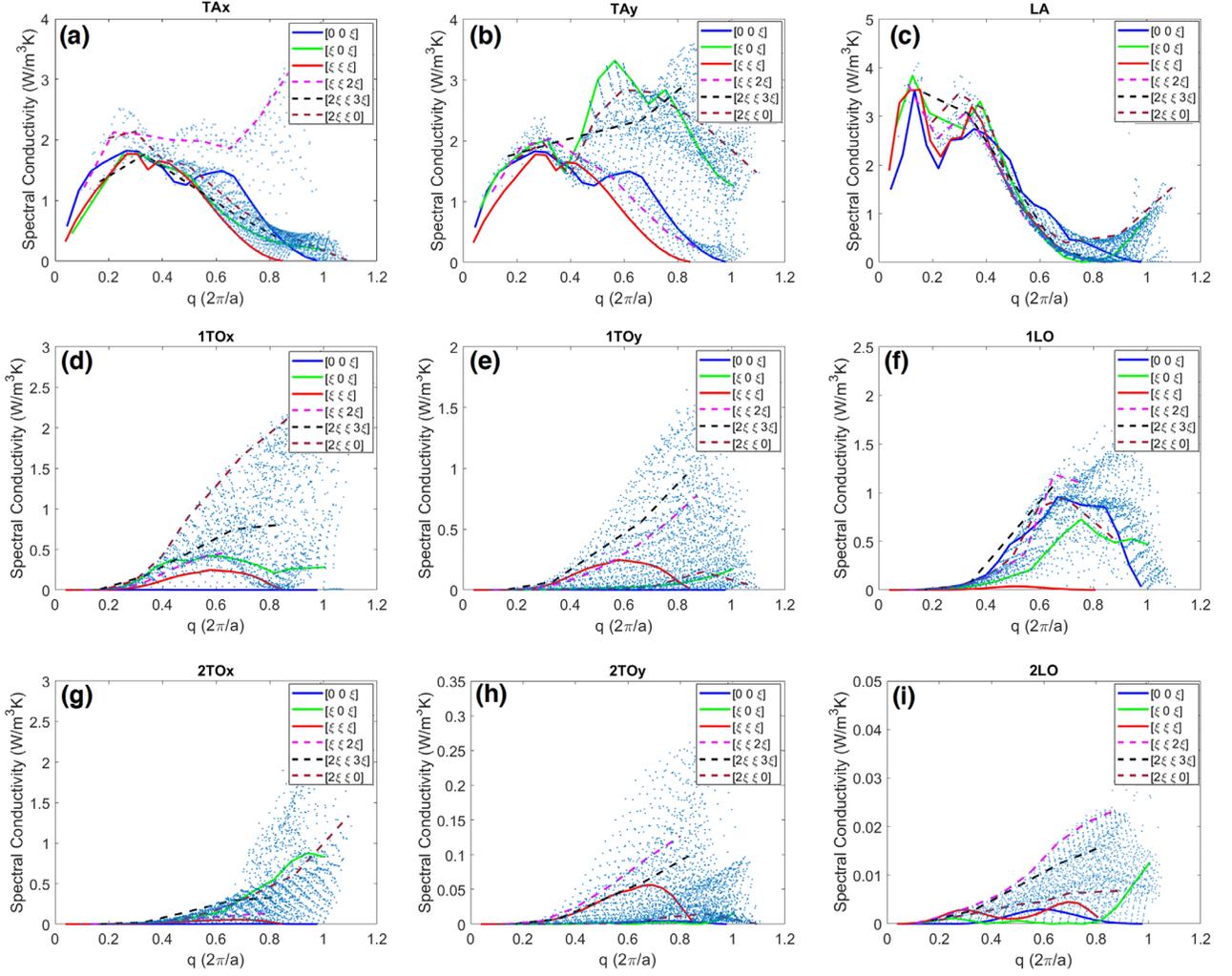

**FIG. S2.** *Spectral Conductivity profile for acoustic a)$TA_x$, b)$TA_y$, c)LA modes and optical d)$1TO_x$ e)$1TO_y$ f)1LO g)$2TO_x$ h)$2TO_y$ i)2LO modes*

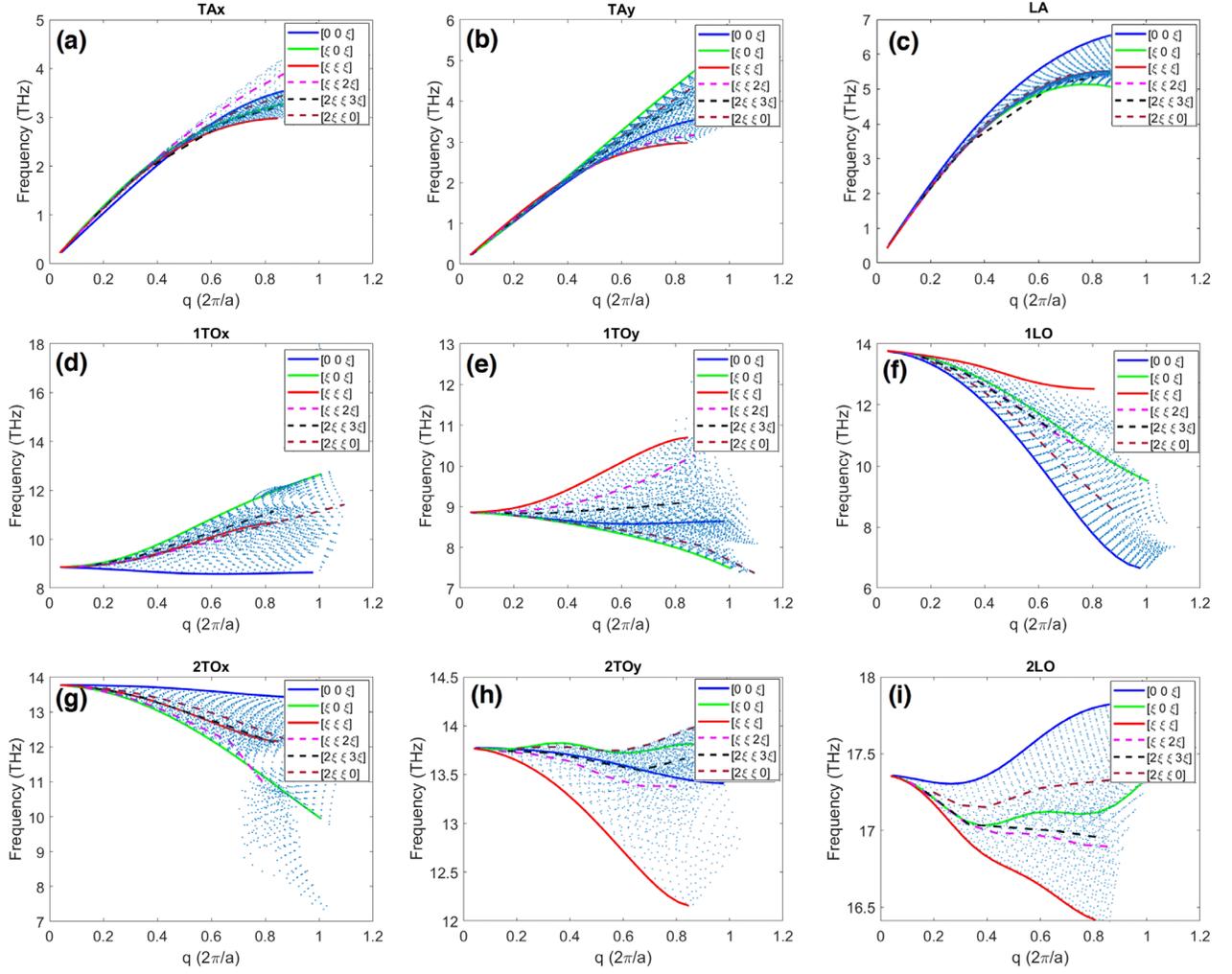

***FIG. S3.*** *Phonon frequency vs q for acoustic a)$TA_x$, b)$TA_y$, c)LA modes and optical d)$1TO_x$ e)$1TO_y$ f)1LO g)$2TO_x$ h)$2TO_y$ i)2LO modes*

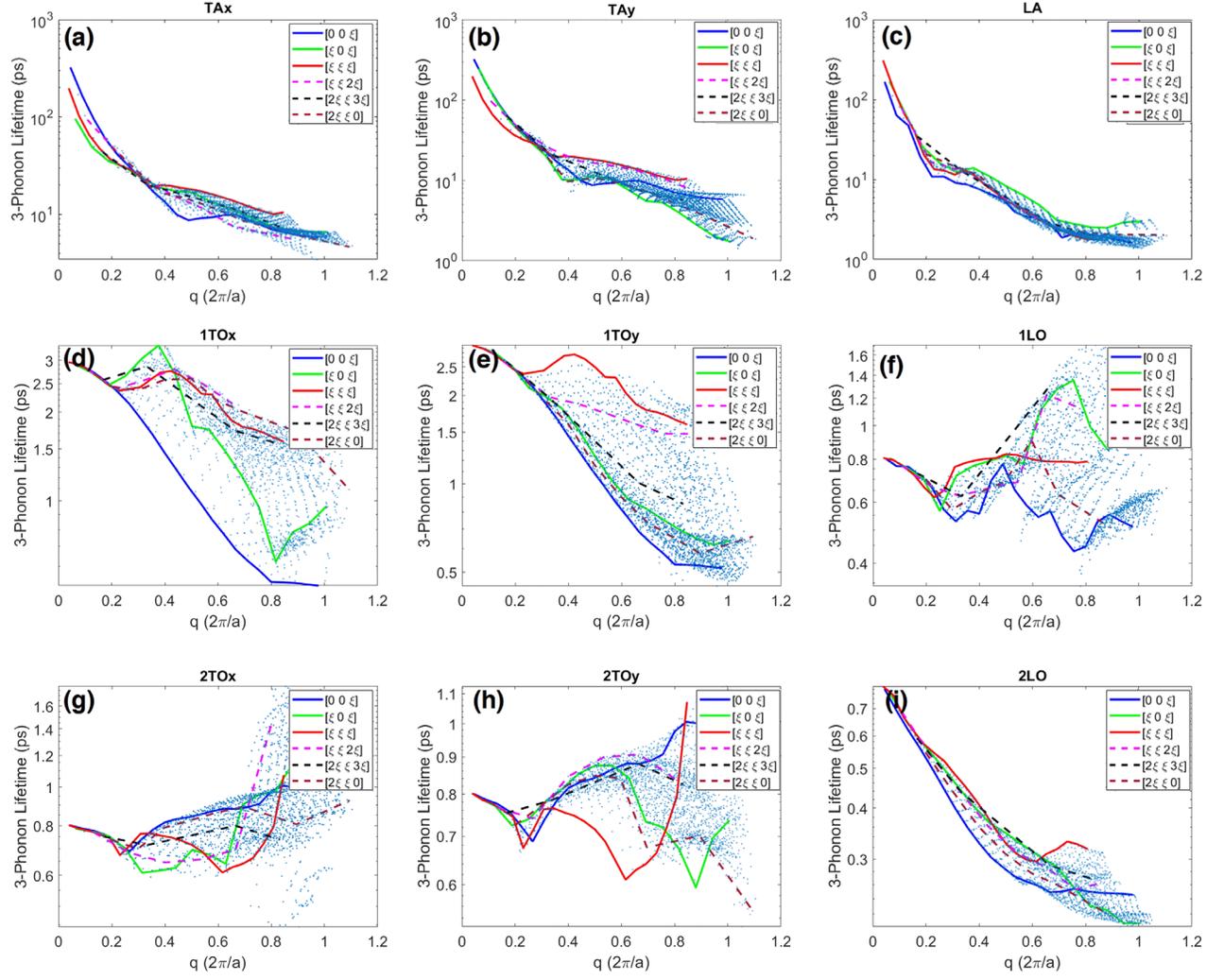

**FIG. S4.** *Phonon lifetimes vs q for acoustic a)$TA_x$, b)$TA_y$, c)LA modes and optical d)$1TO_x$ e)$1TO_y$ f)1LO g)$2TO_x$ h)$2TO_y$ i)2LO modes.*

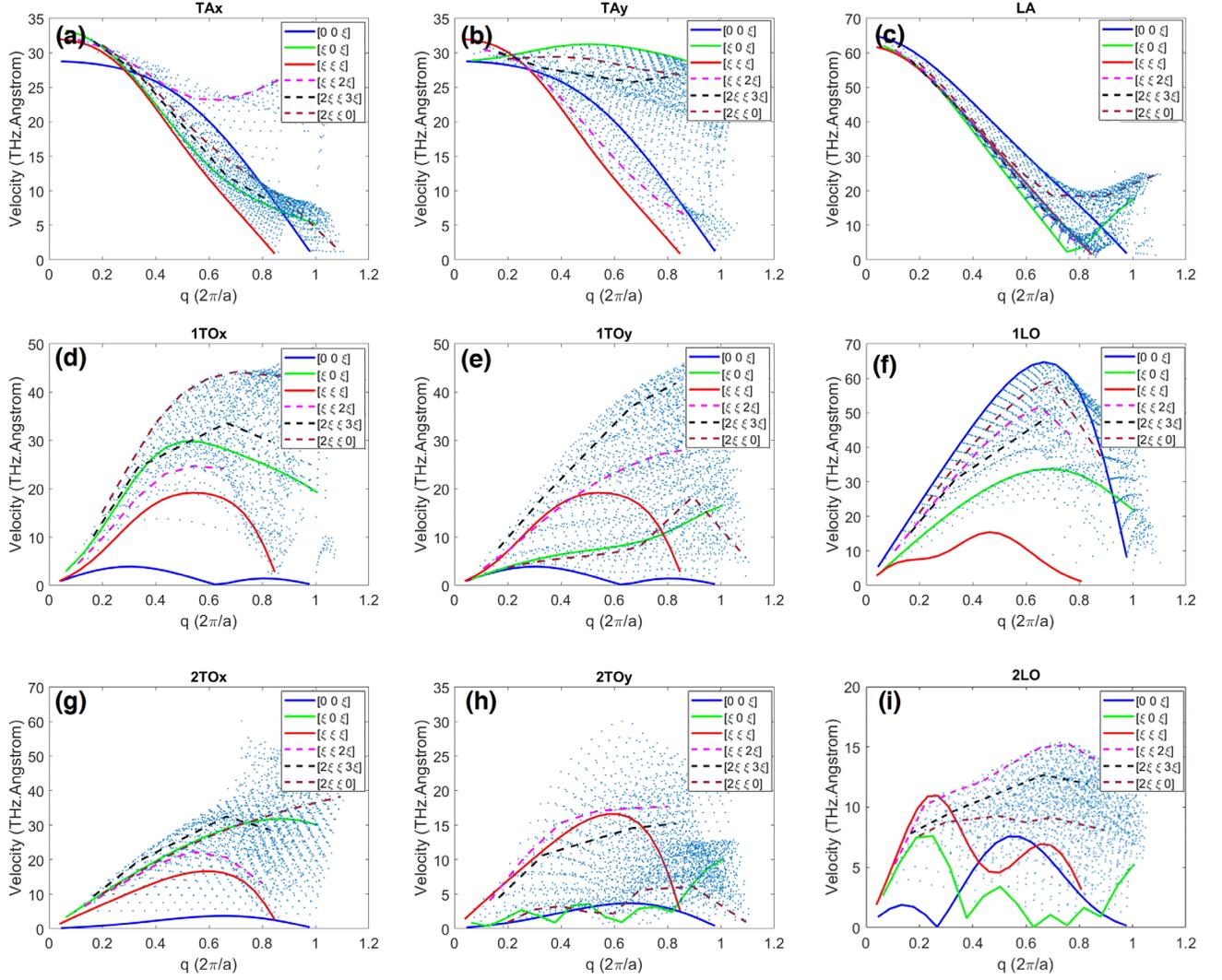

**FIG. S5.** *Phonon velocity vs q for acoustic a)$TA_x$, b)$TA_y$, c)LA modes and optical d)$1TO_x$ e)$1TO_y$ f)1LO g)$2TO_x$ h)$2TO_y$ i)2LO modes.*

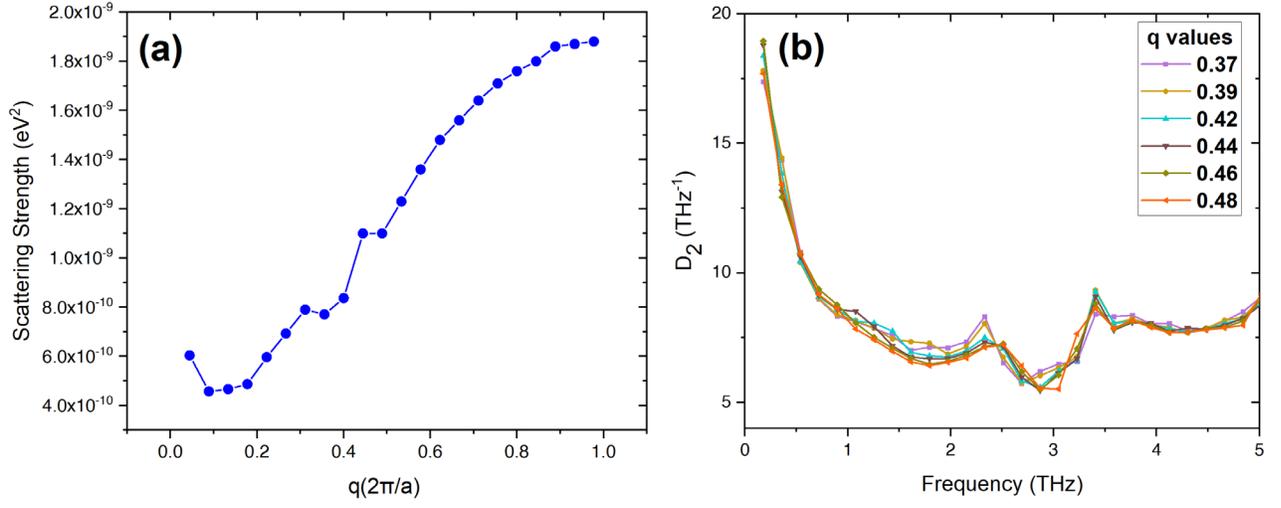

**FIG. S6.** *For the TA mode (a) Scattering Strength of phonons along <100> direction (b) Joint Density of States $D_2$ at different frequency and q points along <100> direction.*

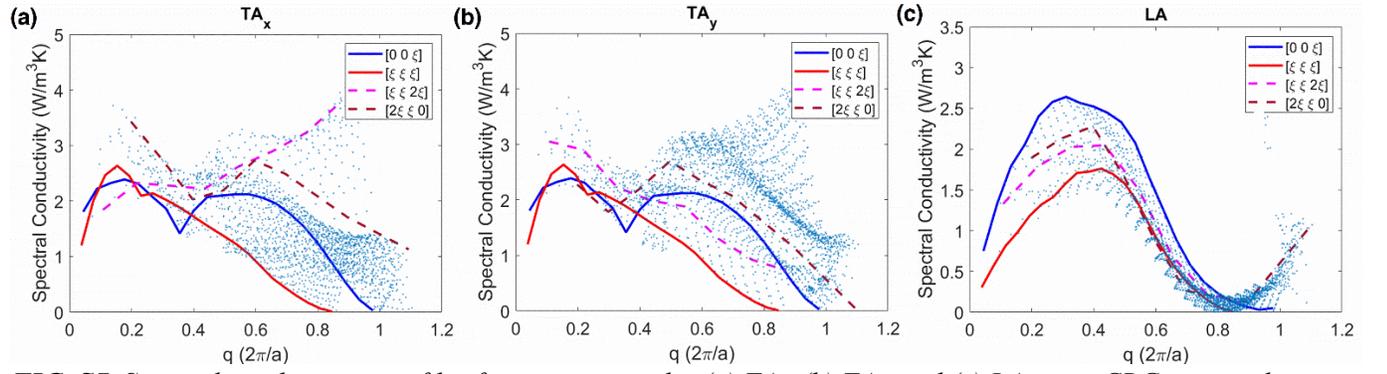

**FIG. S7.** *Spectral conductivity profiles for acoustic modes (a) $TA_x$, (b) $TA_y$, and (c) LA using CRG potential.*

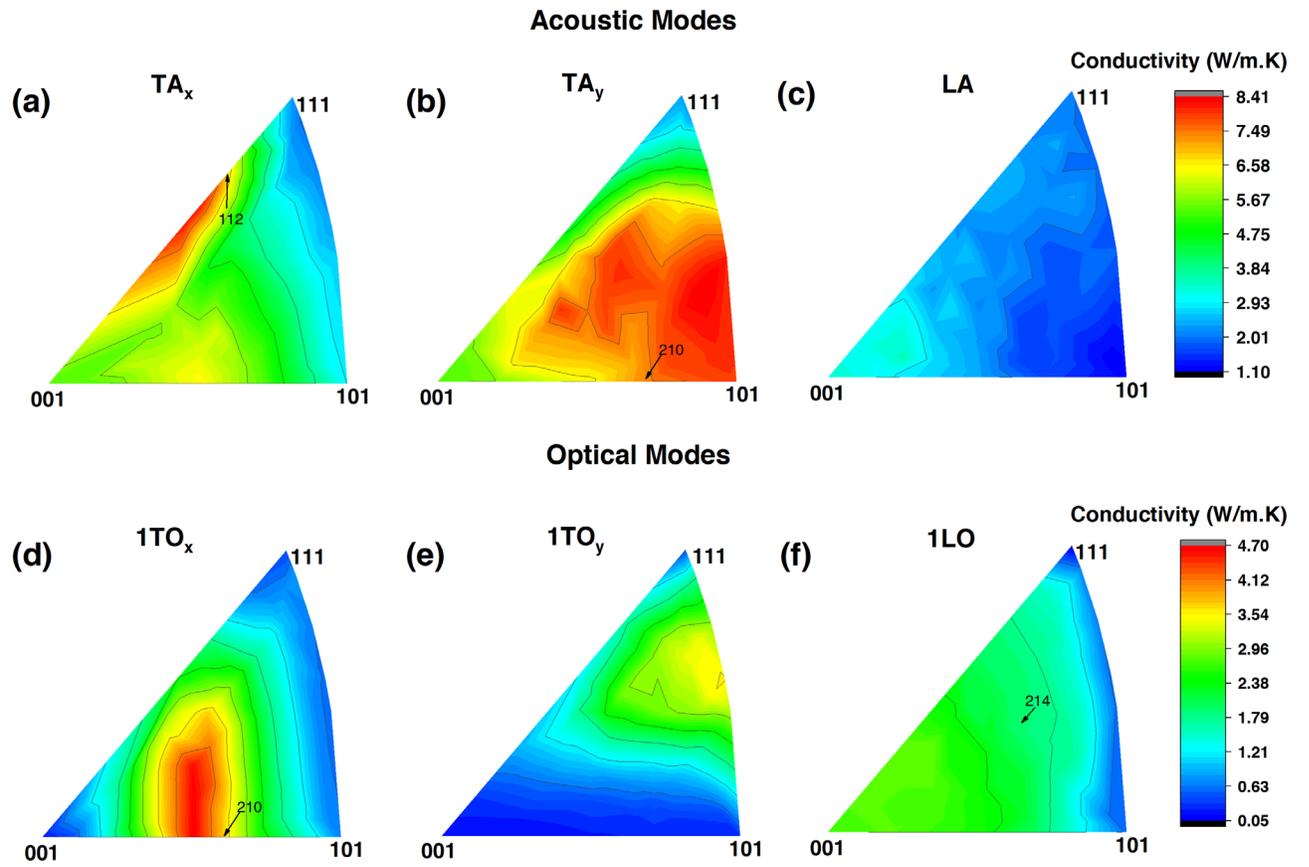

***FIG. S8.*** *Inverse polar plot of the acoustic modes (a) $TA_x$, (b) $TA_y$, (c) LA, and optical modes (d) $1TO_x$, (e) $1TO_y$, (f) 1LO using EIP-CRG at q=0.620.*